\documentclass[twocolumn,showpacs,preprintnumbers,amsmath,amssymb]{revtex4}
\usepackage{graphicx}
\usepackage{dcolumn}
\usepackage{bm}

\begin{document}

\title{Properties of high-$T_c$ copper oxides from band models of spin-phonon coupling.}

\author{T. Jarlborg}

\affiliation{
DPMC, University of Geneva, 24 Quai Ernest-Ansermet, CH-1211 Geneva 4,
Switzerland
}


\begin{abstract}

The mechanism of spin-phonon coupling (SPC) and possible consequencies
for the properties of high-$T_C$ copper oxides are presented.
The results are based on 
ab-initio LMTO band calculations and a nearly free-electron (NFE) model of
the band near $E_F$. Many observed properties are compatible with
SPC, as for the relation between doping and $\vec{q}$ for spin
excitations and their energy dependence. The main pseudogap is
caused by SPC and waves along [1,0,0], but it is suggested that secondary
waves, generated along [1,1,0], contribute to a 'waterfall'
structure. Conditions for optimal $T_C$, and the possiblities
for spin enhancement at the surface are discussed.

\end{abstract}


\maketitle

\section{Introduction.}

The normal state properties of high-$T_c$ copper oxides show many
unusual properties like pseudogaps, stripe-like
charge/spin modulations with particular energy and doping dependencies, Fermi-surface (FS) "arcs" in
the diagonal direction, 'kinks' and 'waterfalls' (WF) in the band dispersions, 
manifestations of isotope shifts, phonon softening
and so on \cite{tran}-\cite{vig}.
Band results for long '1-dimensional' (1-D)
supercells, calculated by the Linear Muffin-Tin Orbital (LMTO)
method
and the local spin-density approximation (LDA), 
show large spin-phonon coupling (SPC)
within the CuO plane of these systems \cite{tj1}. 
This means that an antiferromagnetic (AFM) wave of the correct wave length and the proper phase
is stronger when it coexists with the phonon \cite{tj3}. The LMTO results have been used to parametrize
the strength of potential modulations coming from
phonon distortions and spin waves of different length \cite{tj5}.
These parameters have been used in a nearly free electron (NFE) model in order to
visualize the band effects from the potential modulations in 2-D. Many properties
are consistent with SPC, as have been shown previously \cite{tj1}-\cite{tj7}.

\section{Calculations and Results.}


Ab-initio LMTO band calculations based on the local density approximation (LDA) are made for
La$_{(2-x)}$Ba$_x$CuO$_4$ (LBCO), with the use of the virtual crystal approximation (VCA)
to La-sites to account for doping, $x$.
Calculations for long supercells, mostly oriented along
the CuO bond direction, are used for modeling of phonon distortions
and spin waves \cite{tj5}.
The calculations show that pseudogaps (a 'dip' in the density-of-states, DOS) 
appear at different energies depending
on the wave lengths of the phonon/spin waves. This is consistent with a correlation
between doping and wave length, and with phonon softening in doped systems \cite{tj3}.

The difficulty with ab-initio calculations is that very large
unit cells are needed for realistic 2D-waves. Another shortcoming is that the original
 Brillouin zone is folded by
the use of supercells, which makes interpretation difficult. The band at $E_F$
is free-electron like, with an effective
mass near one, and the potential modulation
and SPC can be studied within the nearly free-electron model (NFE) \cite{tj6}.
The AFM spin arrangement on near-neighbor (NN) Cu along [1,0,0] corresponds to
a potential perturbation, 
$V(\bar{x}) = V_q^t exp(-i\bar{Q} \cdot \bar{x})$ (and correspondingly for $\bar{y}$). 
A further modulation ($\bar{q}$) leads to 1D-stripes perpendicular to $\bar{x}$
(or "checkerboards" in 2-D along $\bar{x}$ and $\bar{y}$), with a modification; 
$V(\bar{x}) 
= V_q^t exp(-i(\bar{Q}-\bar{q}) \cdot \bar{x})$, and the gap moves from the zone boundary to ($\bar{Q}-\bar{q}$)/2. 
\cite{tj6}.

The NFE model reproduce the qualitative results of the full band (1-D) calculation. In 2-D it
leads to a correlation between doping and the amplitude of $V_q^t$,
because the gap (at $E_F$) opens along $(k_x,0)$ and $(0,k_y)$, but not along
the diagonal \cite{tj6}.
The combined effect is that the dip in the total DOS
(at $E_F$)
will not appear at the same band filling for a small and a wide gap.
The $q$ vs. $x$ behavior for a spherical NFE band with $m^*$ close to 1, and parameters $V^t_q$ 
(for one type of phonon) from ref. \cite{tj6} show a saturation, see Fig.1. 
This is quite similar to what is observed \cite{yam}. 
The reason is that no checkerboard solutions are found for larger doping,
but unequal $q_x/2$ (fixed
near 0.11) and $q_y/2$ produce realistic solutions. The DOS at $E_F$, $N$, is lowest
within the pseudogap.
A gap caused by spin waves, disappears at a temperature $T^*$ when
thermal excitations can overcome the gap, and the spin wave can no longer be supported. Therefore,
the spin-waves are most important for the pseudogap (even though
phonons are important via SPC), and $T^*$ is estimated to be
1/4th of the spin part of $V^t_q$ \cite{tj6}.  The opposite $x$-variations of $T^*$ and $N$ 
(note that $\lambda \propto N V_q^2$) provides an argument for optimal conditions for superconductivity
for intermediate $x$ \cite{tj6}. Moreover, the pseudogap competes with superconductivity at underdoping, 
since dynamic SPC would be their common cause. However, there is a possibility to raise $N$ and $T_C$ by  
creating an artificial, static pseudogap through a periodic distribution of dopants or strain. 
Two parameters, the periodicity and the strength of the perturbing potential, 
should be adjusted so that the peak in the DOS above or below
the static pseudogap coincides with $E_F$. 

\begin{figure}
\includegraphics[height=7.0cm,width=8.0cm]{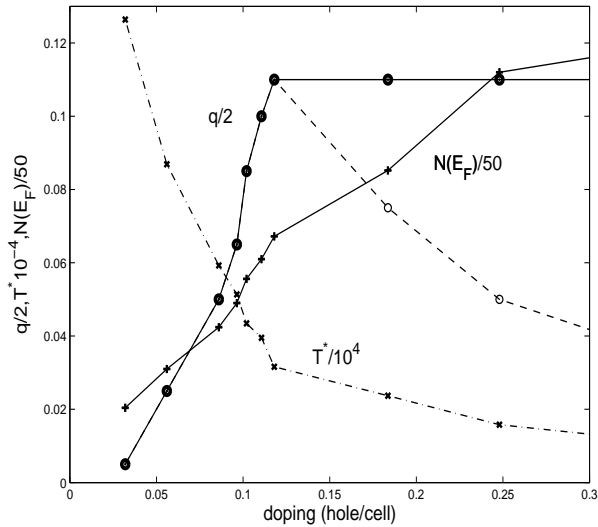}
\caption{The variation of magnetic modulation vector $q/2$ (Circles),
density-of-states in the pseudogap (+-signs $N(E_F)/50$ in states per cell/Ry/spin),
and $T^*/10000$ (x-signs in K), as function of doping $x$ in a NFE model
with parameters from ref. \cite{tj6}. Two different $q$-vectors along
$q_x$ and $q_y$ are needed for doping larger than about 0.12 as indicated by
the filled and open circles.
}
\label{fig0}
\end{figure}

The degree of SPC is different for different phonons.
The total $V_q^t$ with contributions from phonons and spin waves, calculated from LMTO and
information from phonon calculations for Nd$_2$CuO$_4$ \cite{chen},
are 17, 18, 23 and 22 mRy at energies centered around
15 (La), 25 (Cu), 50 (plane-O) and 60 meV (apical-O), respectively \cite{tj7}. 
The results for $x$=0.16 are shown in fig. 2 together with experimental data 
\cite{vig,tran1}. 
The points below 70 meV are for the coupling to
the 4 types of phonons. 
The spectrum is shaped like
an hour-glass with a "waist" at intermediate energy with largest SPC for plane-O.
The solutions for energies larger than 70 meV are independent
of phonons and the exact $(q,\omega)$ behavior is more uncertain \cite{tj7}.
Less doping implies larger $V^t_q$ and longer waves. All $\vec{q}$ become smaller and 
the waist becomes narrower, as can be verified in
LBCO for $x=1/8$ \cite{tran1}, and recently
in lightly doped La$_{1.96}$Sr$_{0.04}$CuO$_4$ \cite{mat}. 
However, the spin modulation in the latter case is in the diagonal direction. 
Heavier O-isotopes will decrease the 
frequencies for the phonons and the coupled spin waves,
and move the waist to lower E.

\begin{figure}
\includegraphics[height=7.0cm,width=8.0cm]{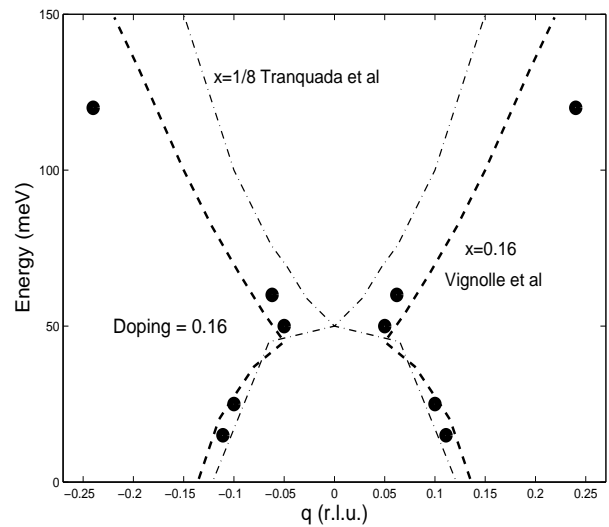}
\caption{Filled circles: Calculated $q-\hbar\omega$ relation from the 2D-NFE model and
the parameters $V_q^t$ for doping
$x=$0.16. The solution without SPC at the largest energy,
is less precise. Broken line: Approximate shape of the
experimental dispersion as it is read from figure 3c in
the work of Vignolle {\it et al} \cite{vig} for La$_{2-x}$Sr$_x$CuO$_4$ at $x=0.16$. 
Thin semi-broken line: Experimental dispersion read from the data by Tranquada {\it et al} \cite{tran1}
on LBCO at lower doping, $x=0.125$.
}
\label{fig1}
\end{figure}

Another odd feature is the WF-dispersion of the band below $E_F$ in the diagonal direction,
seen by ARPES \cite{chang}. The suggestion here is that
this feature comes from a gap below $E_F$ in the diagonal direction.
An inspection of the potential for stripe modulations along [1,0,0]
reveals that the potential becomes modulated also along [1,1,0], albeit in a different fashion. 
The potential
is slowly varying like the absolute value of $sine$-functions with 
different phase along different rows. No NN-AFM potential shifts
are found along [1,1,0], and the dominant Cu-d lobes of wave function for $\vec{k}$ along
[1,1,0] are oriented along [1,1,0] and not along the bond direction.
Various arguments for the effective periodicity, 
partly based on these conditions, indicate that a gap should appear at about 
1/3 of the distance between $\Gamma$ and the $M$ point when the doping is near 0.16.
The effective $V_q$ should be less than half
of the amplitude along [1,0,0]. The result is shown in Fig. 3. 
  The k-position of the
gap and the extreme values of the gap energies ($\sim$ 0.5-1 eV below $E_F$)
are not too far from what is seen experimentally \cite{chang}, but again, the quantitative power
of the NFE-model is limited. It is not clear if the vertical part of the
band dispersion can be observed. A vertical line is connecting
the states above and below the gap in Fig. 3, 
which could be justified for an imperfect gap away from the zone boundary.

\begin{figure}
\includegraphics[height=7.0cm,width=8.0cm]{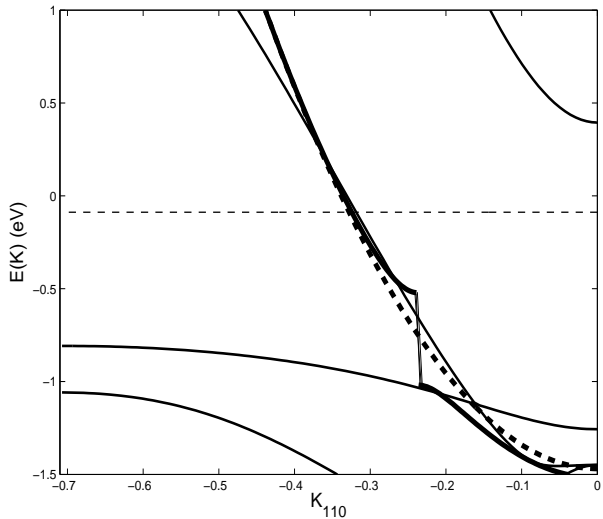}
\caption{Thin lines: LMTO band structure for LBCO between $M$ and $\Gamma$.
Broken line: the FE fit, and the heavy line the NFE solution with a gap.
$E_F$ is at zero in undoped LBCO, and at the thin broken line for $x \sim 0.15$.
}
\label{fig2}
\end{figure}

The dynamics is important if SPC mediates superconductivity. But static, stripe like
features are identified  by
surface tunneling spectroscopy (STM) \cite{davis}. 
 Impurities and defects near the surface
might be important, but also the surface itself could modify the conditions for SPC.   
The latter hypothesis is investigated in LMTO calculations which simulate the surface through
insertion of two layers of empty spheres 
between the outermost LaO-layers. These calculations consider 3 and 5 layers of undoped La$_2$CuO$_4$, and 3 layers
of a doped LBCO with and without phonon distortion in a cell of length 4$a_0$ in the CuO bond direction.
The SPC remains in the surface layer.
The effective doping is in all cases found to increase close to the surface, 
which has 0.1-0.2 more electrons/Cu than
the Cu in the interior,  and
the magnetic moment is 2-3 times larger 
in the surface layer.  The moments disappear without field,
but a calculation for 3 layers of La$_2$CuO$_4$ with a narrower separating layer, has stable AFM
moments $\pm 0.06 \mu_B$ per Cu within the surface layer,
and the local DOS on the Cu at the surface drops near $E_F$.  In addition,
also the apical-O nearest to the surface acquires a sizable moment.
This calculation is simplified, with a probable interaction across the empty layer, but it shows 
that static AFM surface configurations are very close to stability.

\section{Conclusion}

Band calculations show that SPC is important for waves along [1,0,0] or [0,1,0],
with secondary effects in the diagonal direction.
Many properties, like pseudogaps, phonon softening, dynamic stripes, correlation between $\bar{q}$ and $x$,
smearing of the non-diagonal part of the FS, and abrupt disappearance of the spin
fluctuations at a certain $T^*$, are possible consequences of SPC within a
rather conventional band \cite{tj5,tj6}.
Different SPC for different phonons leads to a hour-glass shape of the
$(q,\omega)$-spectrum with the narrowest part for the modes with strongest coupling.
The much discussed WF-structure in the diagonal band dispersion could be a result of
a secondary potential modulation in this direction and a gap below $E_F$.
Static potential modulations within the CuO-planes, such as for superstructures,
could compensate the pseudogap and enhance $N(E_F)$ and $T_C$.
Spin waves become softer through interaction with
phonons and near the surface. These LDA results show a tendency for static spin waves at the surface.

\end{document}